\documentclass[prd, superscriptaddress, nofootinbib]{revtex4}
\usepackage{graphicx, epsfig, color}
\usepackage{amsmath,amsfonts}

\def\eqq#1{Eq.~(\ref{#1})}

\newcommand {\nspec}{N_{\rm spec}}

\newcommand {\fsky}{f_{\rm sky}}

\begin{document}
\title{\Large Catastrophic photometric redshift errors: \\[0.cm]
    weak lensing survey requirements}

\author{Gary Bernstein}
\affiliation{Department of Physics and Astronomy, University of Pennsylvania,
Philadelphia, PA 19104}

\author{Dragan Huterer}
\affiliation{Department of Physics, University of Michigan, 
450 Church St, Ann Arbor, MI 48109-1040}

\begin{abstract}
  We study the sensitivity of weak lensing surveys to the effects of {\it
    catastrophic} redshift errors --- cases where the true redshift is
  misestimated by a significant amount. To compute the biases in cosmological
  parameters, we adopt an efficient linearized analysis where the redshift
  errors are directly related to shifts in the weak lensing convergence power
  spectra. We estimate the number $\nspec$ of unbiased spectroscopic redshifts
  needed to determine the catastrophic error rate well enough that biases in
  cosmological parameters are below statistical errors of weak lensing
  tomography.  While the straightforward estimate of $\nspec$ is $\sim10^6$,
  we find that using only the photometric redshifts with $z\lesssim 2.5$ leads
  to a drastic reduction in $\nspec$ to $\sim 30,000$ while negligibly
  increasing statistical errors in dark energy parameters. Therefore, the size
  of spectroscopic survey needed to control catastrophic errors is similar to
  that previously deemed necessary to constrain the core of the $z_s-z_p$
  distribution. We also study the efficacy of the recent proposal to measure
  redshift errors by cross-correlation between the photo-z and spectroscopic
  samples.  We find that this method requires $\sim10\%$ {\it a priori}
  knowledge of the bias and stochasticity of the outlier population, and is
  also easily confounded by lensing magnification bias.  The cross-correlation
  method is therefore unlikely to supplant the need for a complete
  spectroscopic redshift survey of the source population.
\end{abstract}
\maketitle

\section{Introduction}\label{sec:intro}

Weak gravitational lensing is a very promising cosmological probe that has
potential to accurately map the distribution of dark matter and measure
the properties of dark energy and the neutrino masses (for reviews, see
\cite{Bartelmann_Schneider,Huterer_thesis,Munshi_review,Hoekstra_Jain}).
It is well understood, however, that systematic errors may stand in the way of
weak lensing reaching its full potential --- that is, achieve the {\it
  statistical} errors predicted for future ground and space based surveys such
as the Dark Energy Survey (DES), Large Synoptic Survey Telescope (LSST), and
the Joint Dark Energy Mission (JDEM). Controlling the systematic errors is a
primary concern in these and other surveys so that a variety of dark energy tests
(recently proposed and reviewed by the JDEM Figure of Merit Science
Working Group \cite{FOMSWG}) can be performed to a desired high accuracy. 

Several important sources of systematic errors in weak lensing surveys have
already been studied. Chief among them is the redshift accuracy---approximate,
photometric redshifts are necessary because it is infeasible to obtain optical
spectroscopic redshifts for the huge number ($\sim 10^8$-$10^9$) of galaxies
that future surveys will utilize as lensing sources.  It is therefore
imperative to ensure that statistical errors and systematic biases in the
relation between photometric and spectroscopic redshifts (recently studied in
depth with real data
\cite{Oyaizu07,Banerji07,Lima_zdist,Cunha08,Abdalla_6methods}) do not lead to
appreciable biases in cosmological parameters.

The relation between the photometric and spectroscopic redshift has been
previously modeled as a Gaussian with redshift-dependent bias and scatter. It
is found that both the bias and scatter (that is, quantities $\langle
z_p-z_s\rangle $ and $\langle (z_p-z_s)^2\rangle ^{1/2}$ in each bin of
$\Delta z=0.1$), need to be controlled to about $0.003$ or better in order to
lead to less than $\sim 50\%$ degradation in cosmological parameter accuracies
\cite{Maetal,HTBJ,Kitching_sys}. These constraints are a bit more stringent in
the most general case when the redshift error cannot be described as a
Gaussian \cite{Bernstein_Ma}. These requirements imply that $\nspec\lesssim
10^5$ spectra are required in order to calibrate the photometric redshifts to
the desired accuracy. Gathering such relatively large spectroscopic sample
will be a challenge, setting a limit to the useful depth of weak lensing
surveys.  [While the redshift errors have been well studied, other systematics
are also important, especially theoretical errors in modeling clustering of
galaxies at large and small scales, intrinsic shape alignments, and various
systematic biases that take place during observations
\cite{Huterer_Takada,White_baryons,Hagan_Ma_Kravtsov,Zhan_Knox,Huterer_nulling,
  Zentner_Rudd_Hu,Rudd_Zentner_Kravtsov,
  Shapiro_Cooray,Shapiro_reduced,Heitmann:2004,Heitmann_Coyote1,
  Takada_Bridle,Takada_Jain_NG,Kitching_formfill,
  Joachimi_Schneider,Bernstein_comprehensive,Jarvis_Jain_PSF,Ma_jitter,Guzik_Bernstein,Stabenau_SNAP_sys,
  Heymans_STEP1,Massey_STEP2,PaulinHenriksson_PSF,Amara_Refregier,
  Heavens_align,Croft_align,Crittenden_align,Mackey_align,Jing_align,Heymans_Heavens_align,
  King_Schneider_align,Hirata_SDSS_align,HirataSeljak,Mandelbaum_align,Bridle_Abdalla_align,
  Bridle_King}.]

All of the aforementioned photo-z requirement studies
(e.g.~\cite{HTBJ,Maetal}), however, have modeled the errors as a perturbation
around $z_s-z_p$ relation. While this perturbation was allowed to be large and
to have a nonzero scatter and even skewness (e.g.~\cite{Maetal}), it did not
subsume a general, multimodal error in redshift.

In this paper we would like to remedy this omission by estimating the effect
of {\it catastrophic} redshift errors.  Catastrophic errors are loosely
defined as cases when the photometric redshift is grossly misestimated, i.e.\
when $|z_p-z_s|\sim O(1)$, and are represented by arbitrary ``islands'' in the
$z_p-z_s$ plane.  We develop a formalism that treats these islands as small
``leakages'' (or ``contaminations'') and directly estimates their effect on
bias in cosmological parameters. We then invert the problem by estimating how
many spectroscopic redshifts are required to control catastrophic errors at a
level that makes them harmless for cosmology.

The paper is organized as follows. In \S\ref{sec:formalism} we derive the
relevant equations for the bias in cosmological parameters induced by
misestimated catastrophic redshift errors in a tomographic weak lensing
survey.  In \S\ref{sec:apply_surveys} we apply these methods to a canonical
ambitious weak-lensing cosmology project.  In \S\ref{sec:nspec} we ask: how
large must a {\em complete} spectroscopic redshift survey be in order that the
catastrophic photo-z error rates be measured sufficiently well that remnant
cosmological biases are well below the statistical uncertainties?  Newman
\cite{Newman} has suggested an alternative mode of measuring the photo-z error
distribution, namely the angular cross-correlation of the photometric galaxy
sample nominally at $z_p$ with a spectroscopic sample at $z_s$; in
\S\ref{sec:gal_corr} we investigate whether systematic errors in the photo-z
outlier rates derived from this cross-correlation technique will be small
enough to render cosmological biases insignificant.  The final section
discusses the scaling of these results with critical survey parameters, the
ramifications for survey design, and areas of potential future investigation.

\section{Formalism}\label{sec:formalism}

In this section we establish the formalism that takes us from ``islands'' in
the $z_s-z_p$ plane to biases in cosmological parameters. First, however, we
define the basic observable quantity, the convergence power spectrum, and
its corresponding Fisher information matrix. 

\subsection{Basic observables and the Fisher matrix}

The convergence power spectrum of weak lensing at a fixed multipole $\ell$ and for the $i$th
and $j$th tomographic bin is given by
\begin{equation}
\mathcal{P}_{ij}^{\kappa}(\ell) = {\ell^3\over 2\pi^2} 
\int_0^{\infty} dz \,{W_i(z)\,W_j(z) \over r(z)^2\,H(z)}\,
 \mathcal{P}_{\rm mat}\! \left ({\ell\over r(z)}, z\right ),
\label{eq:pk_l}
\end{equation} 
\noindent where $r(z)$ is the comoving distance, $H(z)$ is the Hubble
parameter, and $\mathcal{P}_{\rm mat}(k, z)$ is the matter power spectrum.
The weights $W_i$ are given, for a flat Universe, by $W_i(\chi) = {3\over 2}\,\Omega_M\,
H_0^2\,g_i(\chi)\, (1+z)$ where $g_i(\chi) = \chi\int_{\chi}^{\infty} d\chi_s
n_i(\chi_s) (\chi_s-\chi)/\chi_s$, $\chi$ is the comoving distance and $n_i$
is the comoving density of galaxies if $\chi_s$ falls in the distance range
bounded by the $i$th redshift bin and zero otherwise.  We employ the redshift
distribution of galaxies of the form $n(z)\propto z^2\exp(-z/z_0)$ that peaks
at $2z_0\simeq 0.9$.

The observed convergence power spectrum is
\begin{equation}
C^{\kappa}_{ij}(\ell)=\mathcal{P}_{ij}^{\kappa}(\ell) + 
\delta_{ij} {\langle \gamma_{\rm int}^2\rangle \over \bar{n}_i},
\label{eq:C_obs}
\end{equation}
\noindent where $\langle\gamma_{\rm int}^2\rangle^{1/2}$ is the rms intrinsic
shear in each component which we assume to be equal to $0.24$,
and $\bar{n}_i$ is the average number of galaxies in the $i$th redshift bin
per steradian.  The cosmological constraints can then be computed from the
Fisher matrix
\begin{equation}
F_{ij} = \sum_{\ell} \,{\partial {\bf C}\over \partial p_i}\,
{\bf Cov}^{-1}\,
{\partial {\bf C}\over \partial p_j},\label{eq:latter_F}
\end{equation}
\noindent where ${\bf Cov}^{-1}$ is the inverse of the covariance matrix 
between the observed power spectra.  For a Gaussian convergence field, its  elements are given by
\begin{equation}
{\rm Cov}\left [C^{\kappa}_{ij}(\ell), C^{\kappa}_{kl}(\ell)\right ] = 
{\delta_{\ell\ell'}\over (2\ell+1)\,f_{\rm sky}\,\Delta \ell}\,
\left [ C^{\kappa}_{ik}(\ell) C^{\kappa}_{jl}(\ell) + 
  C^{\kappa}_{il}(\ell) C^{\kappa}_{jk}(\ell)\right ].
\label{eq:Cov}
\end{equation}
where $\fsky$ is fraction of the sky observed and $\Delta\ell$ is the binning
of the convergence power spectra in multipole. 

Our fiducial SNAP survey described below, without any theoretical systematics, determines
$w_0$ and $w_a$ to accuracies of $\sigma(w_0)=0.089$ and $\sigma(w_a)=0.31$
(corresponding to the pivot value determined to $\sigma(w_p)=0.027$).

\subsection{Biases in the Gaussian limit}
\label{gausslimit}
Consider the general problem
of constraining a vector of cosmological parameters $P=\{p_i\}$ based on
an observed data vector $D=\{D_\alpha\}$.  If the observable quantities
$D_\alpha$ are distributed as Gaussians with covariance matrix $C$, then the
first-order formula for bias in the $i$th parameter, $\Delta p_i$, induced by
a bias $\Delta D$ in the data is (e.g.\
\cite{Knox_Scocc_Dod,Huterer_Turner})
\begin{equation}
\Delta p_i = \sum_{j}(F^{-1})_{ij} \sum_{\alpha\beta} {\partial \bar
  D_\alpha \over \partial p_j} (C^{-1})_{\alpha\beta}\,\Delta D_\beta.
\label{bias1}
\end{equation}
Here $F$ is the Fisher matrix for the cosmological parameters, and is defined
as the second derivative of log likelihood (${\cal L}\equiv-\ln L$) with
respect to the parameters.  The bias above can be more concisely expressed as
\begin{equation}
\Delta P = F^{-1} Q \, \Delta D \equiv  F^{-1} V,
\end{equation}
where we have defined the matrix $Q$ and vector $V$ as
\begin{eqnarray}
Q_{ij}  &\equiv& 
   \sum_{k}{\partial \bar D_k \over \partial p_i} (C^{-1})_{k j}\\[0.2cm]
V &\equiv&  Q\, \Delta D.
\end{eqnarray}

The induced parameter bias is considered unimportant if it is small
compared to the expected statistical variation in the cosmological parameters.  In
the case where the likelihood in the parameter space is Gaussian, 
the likelihood of the bias $\Delta P$ being exceeded by a statistical
fluctuation is determined by
\begin{equation}
\Delta\chi^2 = \Delta P^T F \Delta P = V^T F^{-1} V
\label{dchisq}
\end{equation}

In the Appendices we prove two useful theorems about $\Delta\chi^2$:

\begin{enumerate}

\item The bias significance $\Delta\chi^2$ always {\em decreases} or stays
  fixed when we augment the likelihood with (unbiased) prior information, {\it
    e.g.} data from a non-lensing technique;  

\item $\Delta\chi^2$ always {\em decreases} or stays fixed when we marginalize
  over one or more dimensions of the parameter space.  In the Gaussian limit,
  the bias $\Delta p_i$ is unaffected by marginalization over other
  parameters.
\end{enumerate}

We will use these results later to argue that our requirements on the control
of catastrophic redshift errors are conservative, in the sense that adding
other cosmological data or considering individual cosmological parameter biases
will only weaken the requirements. 

\subsection{The case of catastrophic photo-z errors}

For weak-lensing tomography, the data elements $D_\alpha$ are the
convergence (or shear) cross-power spectrum elements
$C^{\kappa}_{\alpha\beta}(\ell)$ between photo-z bins $\alpha$ and
$\beta$ at multipole $\ell$.  Let us examine how these will be biased
by photo-z outliers.  
[The data covariance matrix $C$ of \S\ref{gausslimit} is the matrix ${\rm Cov}$ of Eq.~(\ref{eq:Cov}).]

We assume a survey with the (true) distribution of source galaxies in
redshift $n_S(z)$, divided into some number $N_b$ of bins
in redshift. Let us define the following terms
\begin{itemize}
\item {\em Leakage:} fraction of objects from a given spectroscopic bin
that are placed into an incorrect (non-corresponding) photometric bin.

\item {\em Contamination:} fraction of galaxies in a given photometric bin
that come from a non-corresponding spectroscopic bin.
\end{itemize}

One could estimate either of these quantities --- after all, when specified
for each bin, they contain the same information.  Let {\em leakage} fraction
$l_{ST}$ of galaxies in some spectroscopic-redshift bin $n_S$ (the ``source''
of leakage) end up in some photo-z bin $n_T$ (the ``target'' of leakage), so
that $l_{ST}$ is the fractional perturbation in the source bin.  Note that,
since bins $S$ and $T$ may not have the same number of galaxies, the
fractional perturbation in the target bin is not the same. The {\em
  contamination} of the target bin $T$, $c_{ST}$, is related to the source bin
leakage via
\begin{equation}
c_{ST} = {N_S\over N_T}\, l_{ST} 
\label{eq:contamination}
\end{equation}

\noindent where $N_S$ and $N_T$ are the absolute galaxy numbers in the source
and target bin respectively. 

\begin{figure}[t]
\includegraphics[scale=0.35]{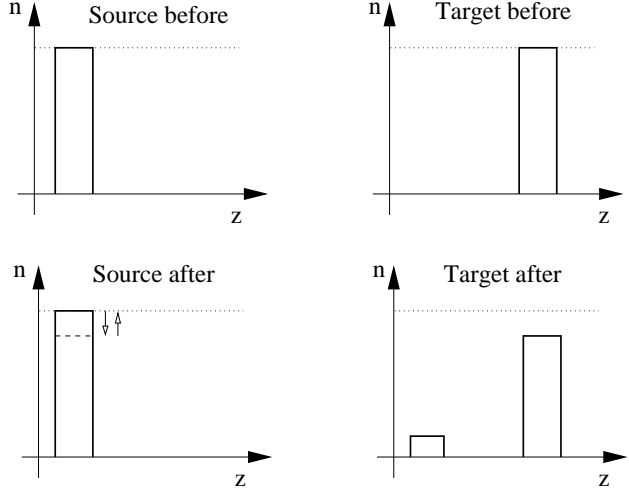}
 \caption{Explanation of how the leakage and contamination operate. In this figure, we
   assume for simplicity that the number of galaxies in the source and
   target bin is the same, so that $l_{ST}=c_{ST}$. Because the redshift distribution
   $n(z)$ is normalized to unit integral in each bin, the source bin's
   redshift distribution $n_S(z)$ does not change; see the bottom left
   panel. The target bin's redshift distribution, $n_T(z)$, does change
   however, as illustrated in the bottom right panel. }
\label{fig:leakage}
\end{figure}

The redshift distribution of galaxies (normalized to unity) in the source bin,
$n_S$, does not change, since a fraction of galaxies is lost --- but the
redshift distribution is normalized to unit integral; see
Fig.~\ref{fig:leakage}.  Conversely, things are perturbed in the target bin,
since it now contains two populations of galaxies, the original one with
fraction $1-c_{ST}$, and the contamination at incorrect (source bin) redshift
with fraction $c_{ST}$; again this is clearly shown in Fig.~\ref{fig:leakage}.
Therefore

\begin{eqnarray}
n_S &\rightarrow & n_S\\[0.1cm]
n_T &\rightarrow & (1-c_{ST})\,n_T + c_{ST}\,n_S
\label{eq:n_bias}
\end{eqnarray}

\noindent and only the {\it target} bin is affected (i.e. biased) by photo-z
catastrophic errors.

The effect on the cross power spectra is now easy to write down. Clearly, only
the $(\alpha, \beta)$ cross-spectra where one of the bins is the
target bin --- $\alpha=T$ or 
$\beta=T$ --- will be affected

\begin{eqnarray}
C_{TT} &\rightarrow &(1-c_{ST})^2C_{TT} +2c_{ST} (1-c_{ST})C_{ST} + c_{ST}^2 C_{SS}
           \nonumber\\[0.1cm]
C_{\alpha T} &\rightarrow &(1-c_{ST})C_{\alpha T} +c_{ST}\, C_{\alpha S} 
           \qquad (\alpha\neq T)
	   \label{eq:C_bias}\\[0.1cm]
C_{\alpha\beta} &\rightarrow &C_{\alpha\beta} 
           \qquad\qquad\qquad\qquad\qquad  ({\rm otherwise})
	   \nonumber
\end{eqnarray}

We have checked that ignoring the quadratic terms in $c_{\alpha\beta}$ leads
to no observable effects to the results (for $c_{ST}=0.001$ contamination).
The biases can now be computed as the right hand side minus the left hand side
in the formulae above. We replace the single index $\alpha$ for data elements
in \eqq{bias1} with the triplet $\ell\alpha\beta$ so that $D_{\ell\alpha\beta}
\equiv C^\kappa_{\alpha\beta}(\ell),$ and we reserve the symbol $C$ for the
covariance of the data elements.  The bias in data induced by the catastrophic
errors is
\begin{equation}
\label{deltal}
\Delta D_{\ell\alpha\beta} = \sum_{\mu\nu} c_{\mu\nu} \left[
\delta_{\alpha\nu}(D_{\ell\mu\beta} - D_{\ell\alpha\beta})
+\delta_{\beta\nu}(D_{\ell\mu\alpha} - D_{\ell\alpha\beta})
\right].
\end{equation}
If we make the further assumption that the convergence is a Gaussian
random field, then we have
\begin{eqnarray}
C_{\ell\alpha\beta,\ell^\prime \gamma\delta} & = &
\delta_{\ell \ell^\prime}\left[D_{\ell\alpha\gamma}D_{\ell\beta\delta}
+ D_{\ell\alpha\delta} D_{\ell\beta\gamma}\right] \\
\Rightarrow \quad(C^{-1})_{\ell\alpha\beta,\ell^\prime \gamma\delta} & = &
{\delta_{\ell \ell^\prime} \over 2} (D_\ell^{-1})_{\alpha\gamma}(D_{\ell}^{-1})_{\beta\delta}.
\label{gausscinv}
\end{eqnarray}
\eqq{bias1} simplifies considerably when we invoke Eqs.~(\ref{deltal}) and
(\ref{gausscinv}):
\begin{eqnarray}
\label{bias2}
\Delta p_i & = & \sum_{j,\mu\nu} (F^{-1})_{ij} M_{j,\mu\nu} c_{\mu\nu},
\\
M_{j,\mu\nu} & \equiv & \sum_\ell \left[
(E^\ell_i)_{\mu\nu}-(E^\ell_i)_{\nu\nu} \right], \\
E^\ell_i & \equiv &  {\partial D_\ell \over \partial p_i} D_\ell^{-1} .
\end{eqnarray}
As a reminder, the Fisher matrix in the case of a zero-mean Gaussian
variable is \cite{TTH}
\begin{equation}
F_{ij} = {1 \over 2} \sum_\ell {\rm Tr}(E^\ell_i E^\ell_j).
\end{equation}

In a cosmological application we will marginalize over all parameters except a
subset of interest $A$.  In the Fisher approximation bias is simply projected
onto the $A$ subset: $\Delta p_A = P_A \Delta p$, 
where $P_A$ is the projection matrix  (see Appendix \ref{margapp}).
If $F_A$ is the marginalized Fisher matrix, then the $\Delta\chi^2$ of the
bias after marginalization is
\begin{equation}
\label{bias3}
\Delta \chi^2 = \sum_{\mu\nu} \sum_{\mu^\prime\nu^\prime} 
c_{\mu\nu} c_{\mu^\prime\nu^\prime} 
\sum_{ij}
M_{i,\mu\nu} \left(F^{-1} P_A^T F_A P_A F^{-1}\right)_{ij} M_{j,\mu^\prime\nu^\prime}.
\end{equation}

\section{Application to canonical surveys}\label{sec:apply_surveys}

For first study we examine a weak lensing survey similar to that
proposed for SNAP \cite{SNAP}, but with the source-galaxy selection
restricted to incur minimal catastrophic error rate.  Evaluation of
other potential surveys could be performed following the same model. 

We take the eight-parameter cosmological model considered by the Dark Energy
Task Force (DETF; \citep{DETF}): dark energy physical density $\Omega_{\rm DE}
h^2$, and equation of state parameters $w_0$ and $w_a$; normalization of the
primordial power spectrum $A$ and spectral index $n$; and matter, baryon, and
curvature physical densities $\Omega_M h^2, \Omega_B h^2,$ and $\Omega_k h^2$.
The fiducial values of these parameters are taken from the 5-year WMAP data
\cite{WMAP5}.  We will assume a Planck CMB prior as specified by the DETF
report.  Recall that application of further priors can only weaken the
requirements on photo-z outliers (see Appendix \ref{addprior_app}).

We assume shear tomography with 20 bins linearly spaced over $0<z<4$ with
$\Delta z=0.2$; we have checked that the results are stable with $\Delta z$.
The redshift distribution and fiducial $c_{\alpha\beta}$ are taken from a
simulation of the photo-z performance of SNAP as described in \cite{Jouvel}.
The procedure is to (1) create a simulated catalog of galaxies; (2) calculate
their noise-free apparent magnitudes in the SNAP photometric bands spanning
the visible and NIR to 1.6~$\mu$m; (3) add the anticipated observational noise
to each magnitude; (4) determine a best-fit galaxy type and redshift with the
template-fitting program {\em Le Phare}\footnote{www.oamp.fr/people/arnouts/LE
  PHARE.html}; (5) examine the 95\% confidence region $z_l<z_p<z_h$ determined
by {\em Le Phare} and retain only those galaxies satisfying \citep{Dahlen}
\begin{equation}
D95 \equiv \ln\left( {1+z_h \over 1+z_l} \right) \le 0.15.
\end{equation}
This strict cut results in a catalog of $\approx 70$ galaxies per arcmin$^2$,
with a median redshift of $z_m\approx 1.2$.  The WL survey is assumed to cover
$\fsky=0.1$ of the full sky, with shape noise of $\sigma_\gamma=0.24$ per
galaxy.  We consider only shear tomography at the 2-point level, as this will
likely maximize the bias imparted by redshift errors.  We also ignore
systematic errors other than redshift outliers, which will likely maximize the
statistical significance of the outlier bias.

We will
henceforth in this paper define a redshift outlier to satisfy
\begin{equation}
\left| \ln{1+z_p \over 1+z_s} \right| > 0.2
\qquad ({\rm catastrophic\,\, outlier\,\, definition}).
\end{equation}
In the simulated photo-z catalog, 2.2\% of the source galaxies are
outliers by this criterion.  In our analyses below we will only
consider biases from photo-z errors meeting this outlier criterion,
i.e.\ we assume the ``core'' of the photo-z distribution is well
determined. 

\begin{figure*}[!th]
\psfig{file=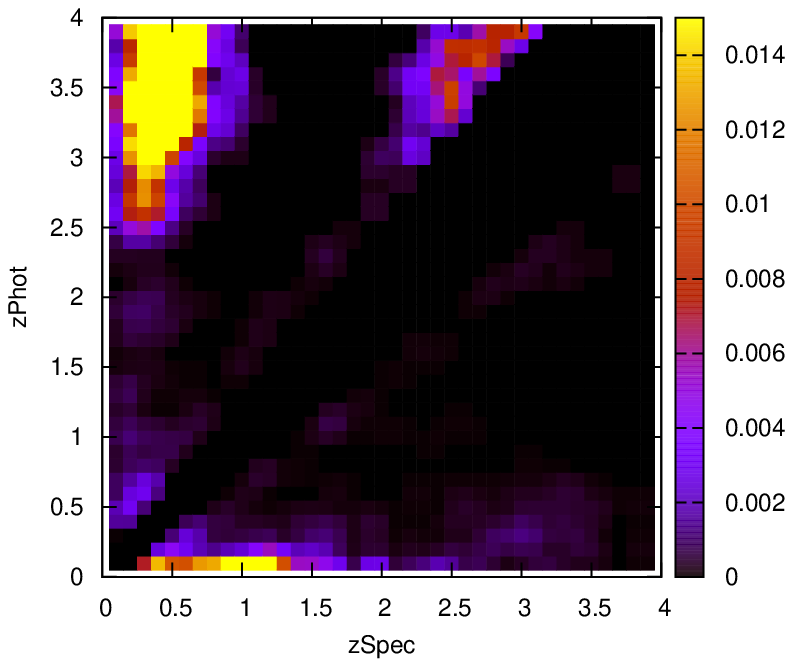,width=3.4in, height=3.2in}
\psfig{file=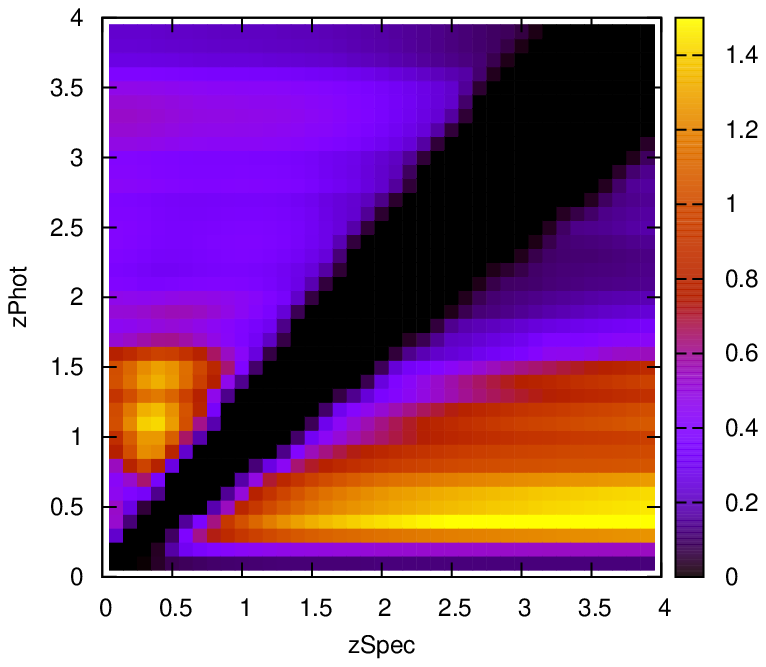,width=3.4in, height=3.2in}
\caption{ {\em Left:} The contamination rate $c_{sp}/\Delta z_s$ of the photo-z
bin per unit redshift in spectro-z is plotted for our example case.
Note that the contamination is highest at $z_p>2.5$ and $z_P<0.2$,
where there are relatively few source galaxies and hence a small
number of outliers can become a large fractional contamination.  {\em
  Right:} The quantity $h_{sp}$ which specifies the significance of the
$w_0-w_a$ bias caused by a contamination rate of $0.001\,\delta z_p$
across a range $\delta z_p$ of photo-z.  This plot indicates that the
outlier contamination rates must be known to 1--3 parts per thousand
over all photometric redshift bins, most sensitively at $0.3<z_p<1.5$.}
\label{fig:zszp1}
\end{figure*}

Figure~\ref{fig:zszp1} illustrates the canonical model and the sensitivity to
redshift outliers in this model.  The left-hand panel shows the quantity
$c_{sp}/\Delta z$ vs $z_s$ and $z_p$. [We scale the contamination by $\Delta
z$ to produce a quantity that is independent of the choice of bin size $\Delta
z$.]  The highest contaminations are in two ``islands'': one at $z_p>2.5$,
$z_s<0.6$ is probably due to confusion between high-$z$ Lyman breaks and
low-$z$ 400-nm breaks.  Because true $z>2.5$ galaxies are relatively rare, a
small leakage rate from $z_s\sim0.5$ can produce a high contamination
fraction.  The {\em Le Phare} code run for this simulation does not incorporate
a magnitude prior for the photo-$z$; doing so might reduce the size of this
island.

A second high-contamination region is $z_s\approx 1$, $z_p<0.2$.  Again the
contamination rate is high because the target-bin density is much lower than
the source-bin density.

The right-hand panel in Figure~\ref{fig:zszp1} shows $\Delta \chi^2$ evaluated
using \eqq{bias3} for this case of catastrophic errors.  We calculate the
significance $\Delta\chi^2_{2d}$ of the bias after marginalization of the
cosmology onto the $w_0-w_a$ plane. We simplify by considering the bias
arising from contamination in a single bin.  This is
\begin{eqnarray}
\Delta\chi^2_{2d} & = & (h_{\mu\nu} \Delta z)^2 c^2_{\mu\nu}, \\[0.2cm]
h^2_{\mu\nu} & \equiv & \sum_{ij}
M_{i,\mu\nu} \left(F^{-1} P_A^T F_A P_A F^{-1}\right)_{ij} M_{j,\mu\nu} / (\Delta z)^2.
\end{eqnarray}
Again the inclusion of the $\Delta z$ factor defines a $h_{\mu\nu}$
that is invariant under rebinning.  The interpretation of $h_{\mu\nu}$
is as follows: if there is an ``island'' of outliers that spans
a range $\delta z_p$ of photo-z bins, and contains a fraction
$\bar c$ of the galaxies in these photo-z bins, 
then the 2d significance of the resultant bias will be
\begin{equation}
\sqrt{\Delta \chi^2} \approx h_{\mu\nu} \delta z_p \bar c.
\label{island1}
\end{equation}
Figure~\ref{fig:zszp1} has been scaled by 1000, so that it indicates the bias
significance of a contamination rate $\bar c=0.001/\delta z_p$.  We desire
$\Delta\chi^2_{2d} \ll 2.3$ to keep the bias well within the 68\% confidence
contour.  The most severe constraint on $\bar c$ would be to take the peak
value $h_{\mu\nu}\approx1300$ and a very wide island, $\delta z_p\approx 0.5$,
in which case the criterion for small bias becomes
\begin{equation}
\bar c < 1/(h_{\mu\nu}\delta z_p) \approx 1/(1300*0.5)=0.0015.
\end{equation}
{\em The contamination rate into any island of outliers
  must be known to 0.0015 or better to avoid significant cosmological
  bias.}
This conclusion is independent of the nominal outlier rate.  The
tolerance on outlier rate will scale with sky coverage as
$\fsky^{-1/2}$.
 
\section{Constraint via spectroscopic sampling}\label{sec:nspec}

The most obvious way to determine the contamination rate
$c_{\alpha\beta}$ is to conduct a {\em complete} spectroscopic
redshift survey of galaxies in photo-z bin $\beta$.  It is of course
essential that the spectra be of sufficient quality to determine
redshifts even for the outliers in the sample.

Let us now estimate the total number of spectra $\nspec$ required in
order to keep the 
total bias below some desired threshold. We will assume that each
redshift drawn from the spectroscopic survey is statistically
independent.  In this case the distribution of $N_{\alpha\beta}$, the
number of galaxies in photo-z bin $\beta$ that have spectro-z in bin
$\alpha$, will be described by a multinomial distribution.  When the
outlier rates are small, the number of spectra in each outlier bin
tend toward independent Poisson distributions.

We would like to relate the contamination uncertainties $\delta
c_{ij}$ to the required number of galaxy spectra. Let $N_\beta$ be the
number of spectra drawn from the photometric redshift bin $\beta$ so
that $\nspec=\sum_\beta N_\beta$. In this case $\langle N_{\alpha\beta}\rangle =
c_{\alpha\beta} N_\beta$ and the variance of the contamination estimate is

\begin{equation}
\delta c^2_{\alpha\beta} = { (\delta N_{\alpha\beta})^2
  \over \langle N_{\beta}\rangle^2} = {c_{\alpha\beta} \over N_\beta }
\label{eq:err_cij}
\end{equation}
Since the Poisson errors between different outlier bins are
uncorrelated, the expected bias significance becomes
\begin{equation}
\langle \Delta \chi^2 \rangle = 
\sum_{\alpha\beta} (h_{\alpha\beta} \Delta z)^2 \langle \delta
c^2_{\alpha\beta}\rangle
 = \sum_{\alpha\beta}  (h_{\alpha\beta} \Delta z)^2
 c_{\alpha\beta}/N_{\beta}.
\label{poisson}
\end{equation}
We would like to quote a total number of spectroscopic redshifts rather than
the number per photo-z slice ($N_\beta$ above) in order to make our
findings more transparent. We consider two cases: first, a slitless or
untargeted spectroscopic survey will obtain redshifts in proportion
to the number density $n_\beta$ of source galaxies in each redshift
bin: $N_\beta = \nspec n_\beta / n$.  Then we will consider a
targeted survey, in which the number $N_\beta$ can be chosen
bin-by-bin to produce the minimal bias for given total $\nspec$.

\subsection{Untargeted spectroscopic survey}\label{sec:untargeted}

\begin{figure}[!t]
\psfig{file=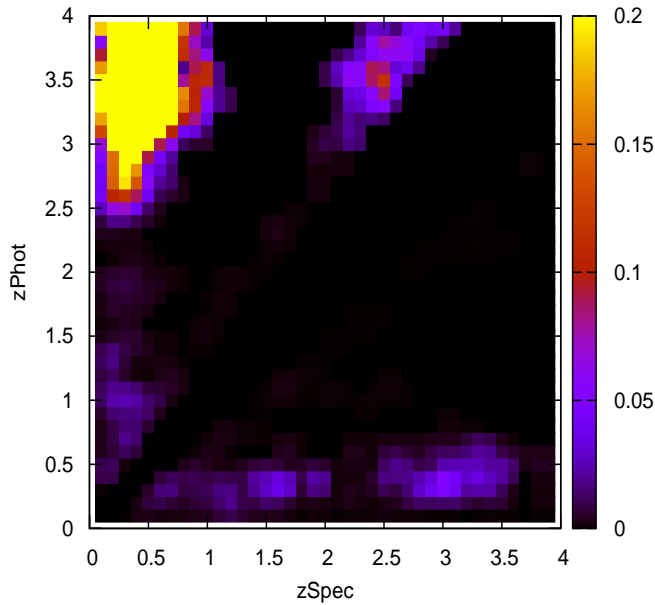,width=3.4in, height=3.2in}
\caption{ Number of calibration spectra required to attain an outlier
  bias significance of   $\Delta\chi^2_{2d}=1$ is given by the
  integral of this plot over the $z_s-z_p$ space.  The quantity
  plotted is, from \eqq{slitless}, $h^2_{sp}c_{sp} n / n_p$, where
  $n_p$ is the source density in the photo-z bin and $n$ is the total
  source density.  The plot is in units of $10^6$ galaxies.
 Note that the required number of calibration
  spectra is strongly driven by the need to constrain the outliers
  with photo-z $z_p>2.5$ but true redshifts $z_s<0.8$.  
}
\label{nshot}
\end{figure}

In the untargeted case, $N_\beta = \nspec n_\beta / n$ and the condition
$\Delta\chi^2 \ll 1$ becomes (from \eqq{poisson})
\begin{equation}
\nspec \gg
\sum_{\alpha\beta} (\Delta z)^2\, h^2_{\alpha\beta} 
 c_{\alpha\beta} n / n_{\beta}.
\label{slitless}
\end{equation}
Figure~\ref{nshot} plots the summand of this expression in the $z_s-z_p$
plane.  The required $\nspec$ is hence the sum over values in this plane (note
that we omit the bins near the diagonal that do not meet the ``outlier''
definition).  We find that $\Delta\chi^2_{2d}\lesssim 1$ requires $\nspec\gtrsim
8\times10^5$, and in the full 8-D parameter space (that is, considering
$\Delta \chi^2$ for the 8-dimensional parameter ellipsoid), we need
$\nspec\gtrsim 2\times10^6$.

These requirements are daunting, potentially larger than the $\nspec$
that are needed to constrain the core of the photo-z distribution as
determined by \citet{Bernstein_Ma}.  Note however that the requirement
is strongly driven by the region $z_s>2.5$, $z_p<0.8$.  This is
because contamination rates are high here in the nominal photo-z
distribution (Figure~\ref{fig:zszp1}), and these bins are sparsely
populated (small $n_p$), meaning that many spectra must be taken in
order to accumulate a strong enough constraint on these contamination
coefficients. 

This suggests a strategy of omitting the $z_p>2.5$ galaxies from the
tomography analysis entirely.  Omitting $z_p>2.5$ from the sum
(\ref{slitless}) produces a much relaxed requirement: for
$\Delta\chi^2_{2d}\lesssim 1$, we need $\nspec\gtrsim 2.8\times10^4$, a
30-fold reduction. This strategy would eliminate $\approx8\%$ of the source
galaxies in the SNAP model, and reduce the dark-energy constraint power by
18\%, as mesured by the DETF figure of merit.  This would be an acceptable
strategy to reduce the outlier bias if one were unable to eliminate the
high-$z_p$ island of outliers by refinements to the photo-z methodology.

\subsection{Targeted spectroscopic survey}\label{sec:targeted}

If we wish to minimize $\Delta \chi^2$ in \eqq{poisson} for a given
total $\nspec$, a simple optimization yields
\begin{eqnarray}
N_\beta & \propto & \sqrt{\sum_\alpha h^2_{\alpha\beta} c_{\alpha\beta}}, \\
\nspec \Delta\chi^2 & = & \left( \sum_\beta \Delta z 
\sqrt{ \sum_\alpha h^2_{\alpha\beta} c_{\alpha\beta}} \right)^2
\label{targeted}
\end{eqnarray}
Optimized targeting reduces the requirement for $\Delta\chi^2_{2d}\lesssim 1$
to be $\nspec\gtrsim 1.2\times10^5$.  Eliminating the $z_p>2.5$ sources
reduces the requirement sixfold, $\nspec\gtrsim 2.0\times10^4$.

Note that the targeted redshift requires $7$ times fewer calibration redshifts
than the untargeted survey, if we are using the full source redshift range,
but only $1.4$ times smaller $\nspec$ if we restrict $z_p<2.5$ in the lensing
analysis.

\subsection{Scaling and Robustness}

The required $\nspec$ to reduce outlier-rate biases to insignificance
scales with the sky coverage and the mean outlier rate as
\begin{equation}
\nspec \propto \fsky \bar c
\end{equation}
when most of the information is coming from shear tomography, and the depth of
the survey is held fixed.  We have verified that $\nspec$ varies little as the
number of tomography bins grows ($\Delta z < 0.2$) and the information content
of the tomography saturates.  The two bias theorems imply that the required
$\nspec$ will drop if we add additional unbiased prior information, or if we
marginalize down to a single dark-energy parameter.

More precisely, we find that the ratio $\nspec\times
(\sigma(w_p)\times\sigma(w_a))\equiv \nspec/{\rm FoM}$ (featuring the well
known ``figure of merit'' \citep{Huterer_Turner,DETF}) is roughly the same
with several alternative survey specifications we consider (and is exactly the
same if only the sky coverage $\fsky$ is varied).

We have used two independent codes to verify the robustness of results to the
myriad of assumptions made and check for the presence of unwanted numerical
artifacts. The two codes agree to roughly a factor of two in $\nspec$, which
is satisfactory given the differences between the implementations, {\em e.g.}
whether curvature and/or neutrino masses are free to vary, and whether the
fiducial redshift distribution is smoothed over the cosmic variance in the
simulated catalog.

\subsection{Correlated outlier errors and incompleteness}

So far we have considered the case where bin-to-bin fluctuations in
contamination $\delta c_{\alpha\beta}$ are uncorrelated. 
Contamination-rate errors $\delta c_{\alpha\beta}$ that are {\it correlated}
from bin to bin might arise if the spectroscopic survey systematically
misses outliers in certain redshifts islands (or if the spectroscopy is not
done at all!).  We can set a specification on the maximum allowable
systematic contamination rate error $\delta \bar c$ in an island of photo-z
width $\delta z_p$ using \eqq{island1}.  Our previous results for the
canonical survey show that the contamination rate in the island should
be known to $\delta \bar c \lesssim 0.0015$.  In other words a
spectroscopic-redshift failure rate of only 0.15\% in some range of
$z_p$ can cause a significant cosmology bias if all of these missed
redshift are outliers in a particular island.  A 99.9\% success rate
has rarely if ever been achieved in a spectroscopic
redshift survey.

\section{Constraining outlier rates using galaxy correlations}\label{sec:gal_corr}

The above requirements on $\nspec$ and completeness may be too expensive to
accomplish, particularly for fainter galaxies.  We now examine the possibility
that one could make use of a spectroscopic galaxy sample that does {\em not}
fairly sample the photo-z galaxies \cite{Newman}.  The idea is to
cross-correlate a photo-z sample at nominal redshift $z_p$ with a
spectroscopic sample known to be confined to a distinct bin $z_s$.  The
amplitude of this cross-correlation will tell us something about the
contamination rate $c_{sp}$, since there is no intrinsic correlation between
the galaxy densities at the two disparate redshifts.

Newman \cite{Newman} calculates the errors in this estimate that would be
induced by shot noise in the sample (for a somewhat related work, see
\cite{Schneider_Knox}).  Here we assume
that statistical errors will be negligible and attend to two systematic errors
that will arise.

\subsection{Outlier bias and correlation coefficients}

First, we define $g_s$, $g_p$ to be the fluctuations in sky density of the
projected distributions of the spectroscopic-survey galaxies in bin $s$ and
the photometric-redshift galaxies in bin $p$.  We set $m_s$ to be the
fractional fluctuations in the projected mass density in redshift bin $s$.
The bias is defined by $\langle g_s^2\rangle = b_s^2\langle m_s^2
\rangle$. The amplitude of the measured cross-correlation between the angular
distributions $g_s$ and $g_p$ of the spectroscopic and photometric samples can
be written as
\begin{equation}
\langle g_s g_p \rangle = c_{sp} {b_s b_{sp} r_{sp} } \langle
m_s^2 \rangle = 
c_{sp} {b_{sp} r_{sp} \over b_s} \langle
g_s^2 \rangle,
\end{equation}
where $b_{sp}$ is the bias of the {\em outlier population}, and $r_{sp}$ is
the correlation coefficient between the density fluctuations of the
spectroscopic and the outlier populations.  The outlier population are those
minority of sources in photo-z bin $p$ that have spectroscopic redshifts in
bin $s$.

In a large survey, shot noise in $\langle g_s g_p \rangle$ and in $ \langle
g_s^2 \rangle$ might become small, and $b_s$ could be determined to good
accuracy through study of the redshift-space power spectrum of the
spectroscopic targets.  It will be difficult, however, to discern $b_{sp}$
because the angular correlation signal of the outlier population is
overwhelmed by the correlations of the galaxies in the ``core'' of the photo-z
error distribution (i.e.\ photo-zs which have {\it not} been catastrophically
misestimated).  The correlation coefficient $r_{sp}$ also has no alternative
observable signature we have identified.  Hence there will be a
systematic-error floor on $\delta c_{sp}$ arising from the finite {\it a
  priori} knowledge of the product $b_{sp} r_{sp}$.

\subsection{Lensing Magnification}
The second complication to the cross-correlation method is that
gravitational lensing magnification bias will induce a correlation
between the spectroscopic and photometric samples even if there is
{\em no}
contamination.  Let us assume that the spectroscopic sample is in the
foreground of the photo-z ``core''; a similar analysis can be done when
using cross-correlation to search for contamination by background
galaxies.  There are two types of magnification-induced correlations.
Following the notation of \cite{HirataSeljak}, there is a ``GG''
correlation, in which both the spectro and photo galaxy samples are
lensed by mass fluctuations in the foreground of both.  Then there is
a ``GI'' effect, in which the mass associated with the fluctuations in
the foreground sample induces magnification bias on the background
sample. 

The GI correlation is as follows: Let the spectroscopic bin $z_s$ span a range
$\Delta\chi_s$ in comoving radial distance.  The matter
fluctuations $m_s$ induce a lensing convergence on the photo-z bin at $z_p$ of
\begin{equation}
\kappa_p = {3 \omega_m \over 2} \Delta\chi_s {\chi_p-\chi_s \over \chi_p} m_s,
\end{equation}
where $\omega_m = \Omega_m h^2=0.127$ is the comoving matter density, 
$\chi_p$ and $\chi_s$ are the comoving angular diameter distances to $z_s$
and $z_p$, and we have assumed a flat Universe.

The lensing magnification will induce apparent density fluctuations in
the background sample as
\begin{equation}
g_p^{\rm lens} = q_p \kappa_p,
\end{equation}
where $q_p$ is a magnification bias factor for the galaxies in the
photo-z bin.  For instance if the selection criteria for the bin were
a simple flux limit, and the intrinsic flux distribution were a power
law $dn/df \propto f^{-s}$, then we would have $q_p=2s-2$.  In general
$q_p$ will be of order unity.

The foreground galaxy distribution $g_s$ has a correlation coefficient $r_s$
with the mass
$m_s$, hence a covariance between populations results:
\begin{eqnarray}
\langle g_s g_p \rangle_{GI} & = & {3 \omega_m \over 2} \Delta\chi_s {\chi_p-\chi_s
  \over \chi_p} q_p b_s r_s \langle m_s^2 \rangle \\
 & = & {3 \omega_m \over 2} \Delta\chi_s {\chi_p-\chi_s
  \over \chi_p} q_p {r_s \over b_s} \langle g_s^2 \rangle
\end{eqnarray}
(we have ignored shot noise in the galaxy auto-correlation).  This
lensing contamination will have to be subtracted from $\langle g_s
g_p\rangle$ in order to extract the information on contamination
$c_{sp}$.  Even if all the cosmological factors are well determined,
the magnification coefficient $q_p$ will have to be empirically
estimated.  Finite accuracy in this estimate will increase $\delta
c_{sp}$.

The GG correlation scales as follows: let $\kappa_s$ be the
convergence induced on the foreground (spectroscopic) sample by mass
at $z<z_s$.  This produces density fluctuations $g_s=q_s\kappa_s$.
This mass induces convergence $\kappa_p \approx \kappa_s r(\chi_s,\chi_p)$
on the background (photo-z) source population, where $r$ is an
integral involving the distributions of foreground mass which must
satisfy $r\ge 1$.  Not concerning ourselves with details, we take $r=1.$
The induced angular correlation will be
\begin{equation}
\langle g_s g_p \rangle_{GG}  = q_s q_p r \langle \kappa_s^2 \rangle.
\end{equation}
Typical RMS values $\kappa_s$ are 0.01--0.02 at cosmological
distances.  The GG lensing correlation must be removed from the signal
to retrieve the contamination fraction, and again even if there is no
shot noise and all distances and lensing amplitudes are known
perfectly, the values of $q_s$ and $q_p$ will only be known to finite
precision. 

\subsection{Estimate of systematic errors}
Summing the GG, GI, and intrinsic contributions, 
the cross-correlation between spectroscopic and photometric samples is
\begin{eqnarray}
\langle g_s g_p \rangle & = & 
\left\{ {3 \omega_m \over 2} \Delta\chi_s {\chi_p-\chi_s
  \over \chi_p} q_p {r_s \over b_s} + c_{sp} {b_{sp} r_{sp} \over b_s}
\right\} 
\langle g_s^2 \rangle + q_s q_p r \langle \kappa_s^2 \rangle \\
\Rightarrow c_{sp} & = & {1 \over b_{sp} r_{sp}} \left[{b_s \langle
    g_s g_p \rangle  \over \langle g_s^2 \rangle } -
{b_s q_s q_p \langle \kappa_s^2\rangle   \over \langle g_s^2 \rangle }
 -   {3 \omega_m \over 2} \Delta\chi_s {\chi_p-\chi_s
  \over \chi_p} r_s q_p \right].
\end{eqnarray}
All of the right-hand quantities are potentially well measured from
the survey data itself or from other cosmological probes, except
the outlier covariance factor $b_{sp}r_{sp}$ and the magnification
coefficients $q_p$ and $q_p$.  Uncertainties in the {\it a priori}
assumed values 
of these factors will propagate into the contamination coefficient as
\begin{equation}
\label{dc2}
(\delta c_{sp})^2 \approx  \left[ \delta(b_{sp}r_{sp})\right]^2 c^2_{sp}
 + \delta q_p^2 \left({3 \omega_m \over 2} \Delta\chi_s {\chi_p-\chi_s
  \over \chi_p} \right)^2
 + \left(\delta q_p^2 + \delta q_s^2\right)
\left({ q \langle \kappa_s^2\rangle   \over \langle g_s^2 \rangle}\right)^2
\end{equation}
Here we assume $b\approx r \approx 1$, $q_s\approx q_p$.  

Earlier we showed that contamination into an outlier ``island'' should be
known to $\delta c_{sp}\le 0.0015$ to avoid significant parameter bias.  Can
such a small contamination be measured using the cross-correlation technique?

\begin{itemize}
\item If the nominal outlier rate is $c_{sp}\approx 10\delta c_{sp}\approx 0.015$, then we require a
  prior estimate of outlier bias/covariance accurate to
  $\delta(b_{sp}r_{sp})<0.1$.  Little will be known about the outlier
  population besides its luminosity range, and the outliers may tend to be
  active galaxies or those with unusual spectra whose clustering properties
  might be deviant as well.  We would consider a 10\% prior knowledge on
  outlier bias to be optimistic but perhaps attainable.

\item For the second (GI) term, if we take the distance factors to be $\approx
  1$, and the outlier population to span a range $\Delta\chi_s\approx 0.3$,
  then the magnification bias coefficient must be known to an accuracy of
  $\delta q_p \lesssim 0.025$.  This accuracy in $q_p$ will be challenging to
  achieve.  If the galaxy selection is by simple magnitude cut, then the slope
  of the counts yields $q_p$ and potentially could be measured to high
  precision.  Weak lensing samples typically have more complex cuts and
  weightings, however, than a simple flux cutoff.  Surface brightness, photo-z accuracy, and
  ellipticity errors are involved, making estimation of $q_p$ more difficult.

\item The third (GG) term places constraints on $\delta q_p$ and
  $\delta q_s$ that will generally be weaker than those from the GI term.
\end{itemize}

If the cross-correlation technique is to determine outlier
contamination fractions to an accuracy that renders them harmless,
then we will need to know the $b_{sp}r_{sp}$ product of the outlier
population to 0.1 or better, and also must know the magnification-bias
coefficients $q$ of our populations to $2\%$ accuracy.   This is true
regardless of sample size, and these tolerances will scale as
$\fsky^{-1/2}$.  The demands on $\delta (b_{sp}r_{sp})$ also becomes
more stringent linearly with the photo-z outlier rate.

We have not considered the possibility of extraneous angular correlations
induced by dust correlated with the foreground galaxy sample, or by dust
in front of both samples (B. Menard, private communication).  This
signal will be present to some degree, though may perhaps be diagnosed
with color information.

In summary, while we have shown here that the cross-correlation technique
proposed by \cite{Newman} is sensitive to catastrophic redshift errors, we
found that prospects of measuring these errors (that is, the contamination
coefficients $c_{sp}$) will be difficult using this technique alone. 

\section{Discussion and Conclusion}\label{sec:conclude}

In this paper we have considered the effects of the previously ignored
catastrophic redshift errors --- cases when the photometric redshift is
grossly misestimated, i.e.\ when $|z_p-z_s|\sim O(1)$, and are represented by
arbitrary ``islands'' in the $z_p-z_s$ plane.  We developed a formalism,
captured by Eqs.~(\ref{eq:C_bias}), that treats these islands as small ``leakages''
(or ``contaminations'') and directly estimates their effect on bias in
cosmological parameters. We then inverted the problem by estimating how many
spectroscopic redshifts are required to control catastrophic errors at a level
that makes them harmless for cosmology. In the process, we have proven two
general-purpose theorems (in the Appendices): that the bias due to systematics
always decreases or stays fixed if 1) (unbiased) prior information is added to
the fiducial survey, or 2) we marginalize over one or more dimensions of the
parameter space.

We found that, at face value, of order million redshifts are required in order
not to bias the dark energy parameter measurements (that is, in order to lead
to $\Delta \chi^2\lesssim 1$ in the $w_0-w_a$ plane). However, the requirement
becomes significantly (30 times) less stringent if we restrict the survey to
redshift $z<2.5$; in that case, $\nspec$ is only of order a few tens of thousands.
Essentially, leakage of galaxies from lower redshift to $z>2.5$ is damaging
since there are few galaxies at such high redshift and the relative bias in
galaxy number is large. Therefore, using only galaxies with $z\lesssim 2.5$
helps dramatically by lowering the required $\nspec$ while degrading the dark
energy (figure-of-merit) constraints by mere $\sim 20$\%.

We have studied two approaches for a spectroscopic survey: the untargeted
one where the number of spectra at each redshift is proportional to the number
of photometric galaxies (\S\ref{sec:untargeted}) and the targeted one
where the number of spectra is optimized to be minimal for a given degradation
in cosmological parameters (\S\ref{sec:targeted}). For the case where galaxies
with $z_p >2.5$ are dropped, the targeted survey gave only a modestly ($\sim
40$\%) smaller required $\nspec$.

We do not imply that these $\nspec$ requirements to apply to all proposed
surveys to high accuracy, although the $O(10^{-3})$ required knowledge on
catastrophe rates is robust.  The calculation should be repeated with the
fiducial photo-z outlier distribution, survey characteristics, and
cosmological parameters of interest to a particular experiment.  

Our work demonstrates for example that efforts to reduce the ``island'' of
catastrophic mis-assignment from $z\approx 0$ to $z\approx 3$, such as
magnitude priors, could greatly reduce the required $\nspec$.  Since
$\nspec\propto \fsky \bar c$ (with $\bar c$ being the mean rate of
catastrophic contamination), it is clear that a photo-z survey with improved
$S/N$ and wavelength coverage to reduce the total catastrophe rate will also
require lower $\nspec$ to calibrate these rates.

Another practical implication of these results is that the spectroscopic
redshift surveys must be of very high completeness---99.9\% if there is a
possibility that all failures could be in an outlier island, but   less if some
fraction of the failures are known to be in the core of the error
distribution.

If an outlier island is known to exist at a particular $(z_s, z_p)$ location,
it may be possible to include the contamination $c_{sp}$ as a free parameter
in the data analysis and marginalize over its value.  It is possible that
self-calibration may reduce the bias in cosmological parameters.  It is likely
infeasible, however, to leave the $c_{sp}$ values over the full $(z_s,z_p)$
plane as free parameters.  We leave self-calibration of outlier rates for
future work.

We have also studied whether the technique proposed by Newman \cite{Newman},
which correlates a photometric sample with a spectroscopic one, can be used to
measure, and thus correct for, catastrophic redshift errors. The advantage of
this approach is that the spectroscopic survey need not be a representative
sampling of the photometric catalog.  While we found that the
cross-correlation technique is sensitive to catastrophic errors (specifically,
the contamination coefficients $c_{sp}$), the contamination coefficient is
degenerate with the value $b_{sp}r_{sp}$ of the bias and stochasticity of the
outlier population.  Furthermore there is a correlation induced by lensing
magnification bias that spoofs the contamination signal.  It will therefore be
difficult to use the cross-correlation technique to constrain outlier rates to
the requisite accuracy.

Overall, we are very optimistic that the catastrophic redshift errors can be
controlled to the desired accuracy. We have identified a simple strategy that
requires only of order 30,000 spectra out to $z\simeq 2.5$ for the calibration
to be successful for a SNAP-type survey. Incidentally, this number of spectra required for the
catastrophic errors is of the same order of magnitude as that required for the
non-catastrophic, ``core'' errors \cite{Maetal,HTBJ,Bernstein_Ma}.  Total
spectroscopic requirements of a survey will be based on the greater of
requirements of these two error regimes.

\section{Acknowledgements}

GB is supported by grant AST-0607667 from the NSF, DOE grant
DE-FG02-95ER40893, and NASA grant BEFS-04-0014-0018.  DH is supported by the
DOE OJI grant under contract DE-FG02-95ER40899, NSF under contract
AST-0807564, and NASA under contract NNX09AC89G.

\appendix
\section{Effect of unbiased priors on bias significance}
\label{addprior_app}
Will a bias get worse or better (more or less significant) when additional
unbiased prior information is added to the likelihood?  It is intuitive that
biases $\Delta P$ should decrease when unbiased information is added. However
$F\rightarrow F+G$ for some new non-negative-definite prior Fisher matrix $G$,
meaning that the statistical errors also shrink.  So which effect wins out?
We prove here that addition of unbiased prior information {\em cannot
  increase} the significance of parameter bias.

The proof is straightforward: \eqq{dchisq} gives the significance
$\Delta\chi^2$ of a bias in terms of the original positive-definite Fisher
matrix $F$ and the vector $V$.  If $G$ is a non-negative-definite prior, then
the change significance of the bias is
\begin{equation}
\Delta\chi^2(\mbox{with prior})
- \Delta\chi^2(\mbox{without prior}) = 
V^T (F+G)^{-1} V - V^T F^{-1} V
\end{equation}
This quantity cannot be positive.  If it were, then there would some
$0<\lambda_0<1$ and a positive-definite matrix $H=F+\lambda_0 G$ such that
\begin{equation}
0 < {\partial \over \partial \lambda} \left[V^T (F + \lambda
  G)^{-1} V\right]_{\lambda_0} = -(H^{-1}V)^T G H^{-1} V.
\end{equation}
If $G$ is non-negative definite, this situation cannot occur.  We
hence conclude that the $\Delta\chi^2$ of some bias is always reduced
(or stays the same) by addition of an unbiased prior.

\section{Effect of marginalization on bias significance}
\label{margapp}
Second we can ask: If we calculate the significance of a bias induced
over a parameter space, then marginalize away parameter vector $B$ to
leave parameter vector $A$, how might the significance differ in the
smaller space?  We show that in the Gaussian limit, {\em
  marginalization always reduces (or leaves unchanged) the
  $\Delta\chi^2$ assigned to the bias, although the $\Delta\chi^2$ per
  DOF may increase.}
To see this: first we note that marginalization over $B$ does not
change the biases in the parameters $A$ if the distribution is
Gaussian.  So the bias in $A$ is simply a projection matrix $P_A$
times $\Delta P$: $\Delta P_A = P_A F^{-1} V$.  The $\Delta\chi^2$
after marginalization down to
the $A$ space is determined by the marginalized Fisher matrix,
$F^\prime = [(F^{-1})_{AA}]^{-1}$.  So we have
\begin{eqnarray}
\nonumber
(\Delta \chi^2)_A & = & V^T F^{-1} P^T_A F^\prime P_A F^{-1} V \\
\label {SMW}
 & = & V^T F^{-1} V - (P_B V)^T (F_{BB})^{-1} (P_B V) \\
 & = & \Delta\chi^2 - (P_B V)^T (F_{BB})^{-1} (P_B V).
\end{eqnarray}
The equivalence in (\ref{SMW}) can be derived from manipulation of the
common expression for the inverse of a matrix decomposed into an
$2\times2$ array of submatrices.  Because $F_{BB}$ and its inverse
must be non-negative-definite, the last term is negative, so we are
assured that $(\Delta \chi^2)_A \le \Delta\chi^2$.  Equality is,
however, easily obtained, for example if there is no bias in the $B$
parameters.  We thus know that $\Delta\chi^2 / N_{\rm DOF}$ can
potentially increase.

\bibliography{catastrophic}

\begin{thebibliography}{60}
\expandafter\ifx\csname natexlab\endcsname\relax\def\natexlab#1{#1}\fi
\expandafter\ifx\csname bibnamefont\endcsname\relax
  \def\bibnamefont#1{#1}\fi
\expandafter\ifx\csname bibfnamefont\endcsname\relax
  \def\bibfnamefont#1{#1}\fi
\expandafter\ifx\csname citenamefont\endcsname\relax
  \def\citenamefont#1{#1}\fi
\expandafter\ifx\csname url\endcsname\relax
  \def\url#1{\texttt{#1}}\fi
\expandafter\ifx\csname urlprefix\endcsname\relax\def\urlprefix{URL }\fi
\providecommand{\bibinfo}[2]{#2}
\providecommand{\eprint}[2][]{\url{#2}}

\bibitem[{\citenamefont{Bartelmann and Schneider}(2001)}]{Bartelmann_Schneider}
\bibinfo{author}{\bibfnamefont{M.}~\bibnamefont{Bartelmann}} \bibnamefont{and}
  \bibinfo{author}{\bibfnamefont{P.}~\bibnamefont{Schneider}},
  \bibinfo{journal}{Phys. Rept.} \textbf{\bibinfo{volume}{340}},
  \bibinfo{pages}{291} (\bibinfo{year}{2001}), \eprint{astro-ph/9912508}.

\bibitem[{\citenamefont{Huterer}(2002)}]{Huterer_thesis}
\bibinfo{author}{\bibfnamefont{D.}~\bibnamefont{Huterer}},
  \bibinfo{journal}{Phys. Rev.} \textbf{\bibinfo{volume}{D65}},
  \bibinfo{pages}{063001} (\bibinfo{year}{2002}), \eprint{astro-ph/0106399}.

\bibitem[{\citenamefont{Munshi et~al.}(2008)\citenamefont{Munshi, Valageas,
  Van~Waerbeke, and Heavens}}]{Munshi_review}
\bibinfo{author}{\bibfnamefont{D.}~\bibnamefont{Munshi}},
  \bibinfo{author}{\bibfnamefont{P.}~\bibnamefont{Valageas}},
  \bibinfo{author}{\bibfnamefont{L.}~\bibnamefont{Van~Waerbeke}},
  \bibnamefont{and} \bibinfo{author}{\bibfnamefont{A.}~\bibnamefont{Heavens}},
  \bibinfo{journal}{Phys. Rept.} \textbf{\bibinfo{volume}{462}},
  \bibinfo{pages}{67} (\bibinfo{year}{2008}), \eprint{astro-ph/0612667}.

\bibitem[{\citenamefont{Hoekstra and Jain}(2008)}]{Hoekstra_Jain}
\bibinfo{author}{\bibfnamefont{H.}~\bibnamefont{Hoekstra}} \bibnamefont{and}
  \bibinfo{author}{\bibfnamefont{B.}~\bibnamefont{Jain}}
  (\bibinfo{year}{2008}), \eprint{arXiv:0805.0139}.

\bibitem[{\citenamefont{Albrecht et~al.}(2009)}]{FOMSWG}
\bibinfo{author}{\bibfnamefont{A.}~\bibnamefont{Albrecht}} \bibnamefont{et~al.}
  (\bibinfo{year}{2009}), \eprint{arXiv:0901.0721}.

\bibitem[{\citenamefont{Oyaizu et~al.}(2007)\citenamefont{Oyaizu, Lima, Cunha,
  Lin, and Frieman}}]{Oyaizu07}
\bibinfo{author}{\bibfnamefont{H.}~\bibnamefont{Oyaizu}},
  \bibinfo{author}{\bibfnamefont{M.}~\bibnamefont{Lima}},
  \bibinfo{author}{\bibfnamefont{C.~E.} \bibnamefont{Cunha}},
  \bibinfo{author}{\bibfnamefont{H.}~\bibnamefont{Lin}}, \bibnamefont{and}
  \bibinfo{author}{\bibfnamefont{J.}~\bibnamefont{Frieman}}
  (\bibinfo{year}{2007}), \eprint{arXiv:0711.0962}.

\bibitem[{\citenamefont{Banerji et~al.}(2007)\citenamefont{Banerji, Abdalla,
  Lahav, and Lin}}]{Banerji07}
\bibinfo{author}{\bibfnamefont{M.}~\bibnamefont{Banerji}},
  \bibinfo{author}{\bibfnamefont{F.~B.} \bibnamefont{Abdalla}},
  \bibinfo{author}{\bibfnamefont{O.}~\bibnamefont{Lahav}}, \bibnamefont{and}
  \bibinfo{author}{\bibfnamefont{H.}~\bibnamefont{Lin}} (\bibinfo{year}{2007}),
  \eprint{arXiv:0711.1059}.

\bibitem[{\citenamefont{Lima et~al.}(2008)}]{Lima_zdist}
\bibinfo{author}{\bibfnamefont{M.}~\bibnamefont{Lima}} \bibnamefont{et~al.},
  \bibinfo{journal}{Mon. Not. Roy. Astron. Soc.}
  \textbf{\bibinfo{volume}{390}}, \bibinfo{pages}{118} (\bibinfo{year}{2008}),
  \eprint{arXiv:0801.3822}.

\bibitem[{\citenamefont{Cunha et~al.}(2008)\citenamefont{Cunha, Lima, Oyaizu,
  Frieman, and Lin}}]{Cunha08}
\bibinfo{author}{\bibfnamefont{C.~E.} \bibnamefont{Cunha}},
  \bibinfo{author}{\bibfnamefont{M.}~\bibnamefont{Lima}},
  \bibinfo{author}{\bibfnamefont{H.}~\bibnamefont{Oyaizu}},
  \bibinfo{author}{\bibfnamefont{J.}~\bibnamefont{Frieman}}, \bibnamefont{and}
  \bibinfo{author}{\bibfnamefont{H.}~\bibnamefont{Lin}} (\bibinfo{year}{2008}),
  \eprint{arXiv:0810.2991}.

\bibitem[{\citenamefont{Abdalla et~al.}(2008)\citenamefont{Abdalla, Banerji,
  Lahav, and Rashkov}}]{Abdalla_6methods}
\bibinfo{author}{\bibfnamefont{F.~B.} \bibnamefont{Abdalla}},
  \bibinfo{author}{\bibfnamefont{M.}~\bibnamefont{Banerji}},
  \bibinfo{author}{\bibfnamefont{O.}~\bibnamefont{Lahav}}, \bibnamefont{and}
  \bibinfo{author}{\bibfnamefont{V.}~\bibnamefont{Rashkov}}
  (\bibinfo{year}{2008}), \eprint{arXiv:0812.3831}.

\bibitem[{\citenamefont{Ma et~al.}(2005)\citenamefont{Ma, Hu, and
  Huterer}}]{Maetal}
\bibinfo{author}{\bibfnamefont{Z.-M.} \bibnamefont{Ma}},
  \bibinfo{author}{\bibfnamefont{W.}~\bibnamefont{Hu}}, \bibnamefont{and}
  \bibinfo{author}{\bibfnamefont{D.}~\bibnamefont{Huterer}},
  \bibinfo{journal}{Astrophys. J.} \textbf{\bibinfo{volume}{636}},
  \bibinfo{pages}{21} (\bibinfo{year}{2005}), \eprint{astro-ph/0506614}.

\bibitem[{\citenamefont{Huterer et~al.}(2006)\citenamefont{Huterer, Takada,
  Bernstein, and Jain}}]{HTBJ}
\bibinfo{author}{\bibfnamefont{D.}~\bibnamefont{Huterer}},
  \bibinfo{author}{\bibfnamefont{M.}~\bibnamefont{Takada}},
  \bibinfo{author}{\bibfnamefont{G.}~\bibnamefont{Bernstein}},
  \bibnamefont{and} \bibinfo{author}{\bibfnamefont{B.}~\bibnamefont{Jain}},
  \bibinfo{journal}{Mon. Not. Roy. Astron. Soc.}
  \textbf{\bibinfo{volume}{366}}, \bibinfo{pages}{101} (\bibinfo{year}{2006}),
  \eprint{astro-ph/0506030}.

\bibitem[{\citenamefont{Kitching
  et~al.}(2008{\natexlab{a}})\citenamefont{Kitching, Taylor, and
  Heavens}}]{Kitching_sys}
\bibinfo{author}{\bibfnamefont{T.~D.} \bibnamefont{Kitching}},
  \bibinfo{author}{\bibfnamefont{A.~N.} \bibnamefont{Taylor}},
  \bibnamefont{and} \bibinfo{author}{\bibfnamefont{A.~F.}
  \bibnamefont{Heavens}} (\bibinfo{year}{2008}{\natexlab{a}}),
  \eprint{arXiv:801.3270}.

\bibitem[{\citenamefont{Ma and Bernstein}(2008)}]{Bernstein_Ma}
\bibinfo{author}{\bibfnamefont{Z.}~\bibnamefont{Ma}} \bibnamefont{and}
  \bibinfo{author}{\bibfnamefont{G.}~\bibnamefont{Bernstein}},
  \bibinfo{journal}{Astrophys. J.} \textbf{\bibinfo{volume}{682}},
  \bibinfo{pages}{39} (\bibinfo{year}{2008}), \eprint{arXiv:0712.1562}.

\bibitem[{\citenamefont{Huterer and Takada}(2005)}]{Huterer_Takada}
\bibinfo{author}{\bibfnamefont{D.}~\bibnamefont{Huterer}} \bibnamefont{and}
  \bibinfo{author}{\bibfnamefont{M.}~\bibnamefont{Takada}},
  \bibinfo{journal}{Astropart. Phys.} \textbf{\bibinfo{volume}{23}},
  \bibinfo{pages}{369} (\bibinfo{year}{2005}), \eprint{astro-ph/0412142}.

\bibitem[{\citenamefont{Heitmann et~al.}(2005)\citenamefont{Heitmann, Ricker,
  Warren, and Habib}}]{Heitmann:2004}
\bibinfo{author}{\bibfnamefont{K.}~\bibnamefont{Heitmann}},
  \bibinfo{author}{\bibfnamefont{P.~M.} \bibnamefont{Ricker}},
  \bibinfo{author}{\bibfnamefont{M.~S.} \bibnamefont{Warren}},
  \bibnamefont{and} \bibinfo{author}{\bibfnamefont{S.}~\bibnamefont{Habib}},
  \bibinfo{journal}{Astrophys. J. Suppl.} \textbf{\bibinfo{volume}{160}},
  \bibinfo{pages}{28} (\bibinfo{year}{2005}), \eprint{astro-ph/0411795}.

\bibitem[{\citenamefont{Heitmann et~al.}(2008)\citenamefont{Heitmann, White,
  Wagner, Habib, and Higdon}}]{Heitmann_Coyote1}
\bibinfo{author}{\bibfnamefont{K.}~\bibnamefont{Heitmann}},
  \bibinfo{author}{\bibfnamefont{M.}~\bibnamefont{White}},
  \bibinfo{author}{\bibfnamefont{C.}~\bibnamefont{Wagner}},
  \bibinfo{author}{\bibfnamefont{S.}~\bibnamefont{Habib}}, \bibnamefont{and}
  \bibinfo{author}{\bibfnamefont{D.}~\bibnamefont{Higdon}}
  (\bibinfo{year}{2008}), \eprint{arXiv:0812.1052}.

\bibitem[{\citenamefont{Zentner et~al.}(2008)\citenamefont{Zentner, Rudd, and
  Hu}}]{Zentner_Rudd_Hu}
\bibinfo{author}{\bibfnamefont{A.~R.} \bibnamefont{Zentner}},
  \bibinfo{author}{\bibfnamefont{D.~H.} \bibnamefont{Rudd}}, \bibnamefont{and}
  \bibinfo{author}{\bibfnamefont{W.}~\bibnamefont{Hu}}, \bibinfo{journal}{Phys.
  Rev.} \textbf{\bibinfo{volume}{D77}}, \bibinfo{pages}{043507}
  (\bibinfo{year}{2008}), \eprint{arXiv:0709.4029}.

\bibitem[{\citenamefont{Takada and Jain}(2008)}]{Takada_Jain_NG}
\bibinfo{author}{\bibfnamefont{M.}~\bibnamefont{Takada}} \bibnamefont{and}
  \bibinfo{author}{\bibfnamefont{B.}~\bibnamefont{Jain}}
  (\bibinfo{year}{2008}), \eprint{arXiv:0810.4170}.

\bibitem[{\citenamefont{White}(2004)}]{White_baryons}
\bibinfo{author}{\bibfnamefont{M.~J.} \bibnamefont{White}},
  \bibinfo{journal}{Astropart. Phys.} \textbf{\bibinfo{volume}{22}},
  \bibinfo{pages}{211} (\bibinfo{year}{2004}), \eprint{astro-ph/0405593}.

\bibitem[{\citenamefont{Hagan et~al.}(2005)\citenamefont{Hagan, Ma, and
  Kravtsov}}]{Hagan_Ma_Kravtsov}
\bibinfo{author}{\bibfnamefont{B.}~\bibnamefont{Hagan}},
  \bibinfo{author}{\bibfnamefont{C.-P.} \bibnamefont{Ma}}, \bibnamefont{and}
  \bibinfo{author}{\bibfnamefont{A.~V.} \bibnamefont{Kravtsov}},
  \bibinfo{journal}{Astrophys. J.} \textbf{\bibinfo{volume}{633}},
  \bibinfo{pages}{537} (\bibinfo{year}{2005}), \eprint{astro-ph/0504557}.

\bibitem[{\citenamefont{Zhan and Knox}(2004)}]{Zhan_Knox}
\bibinfo{author}{\bibfnamefont{H.}~\bibnamefont{Zhan}} \bibnamefont{and}
  \bibinfo{author}{\bibfnamefont{L.}~\bibnamefont{Knox}},
  \bibinfo{journal}{Astrophys. J.} \textbf{\bibinfo{volume}{616}},
  \bibinfo{pages}{L75} (\bibinfo{year}{2004}), \eprint{astro-ph/0409198}.

\bibitem[{\citenamefont{Huterer and White}(2005)}]{Huterer_nulling}
\bibinfo{author}{\bibfnamefont{D.}~\bibnamefont{Huterer}} \bibnamefont{and}
  \bibinfo{author}{\bibfnamefont{M.~J.} \bibnamefont{White}},
  \bibinfo{journal}{Phys. Rev.} \textbf{\bibinfo{volume}{D72}},
  \bibinfo{pages}{043002} (\bibinfo{year}{2005}), \eprint{astro-ph/0501451}.

\bibitem[{\citenamefont{Shapiro and Cooray}(2006)}]{Shapiro_Cooray}
\bibinfo{author}{\bibfnamefont{C.}~\bibnamefont{Shapiro}} \bibnamefont{and}
  \bibinfo{author}{\bibfnamefont{A.}~\bibnamefont{Cooray}},
  \bibinfo{journal}{JCAP} \textbf{\bibinfo{volume}{0603}}, \bibinfo{pages}{007}
  (\bibinfo{year}{2006}), \eprint{astro-ph/0601226}.

\bibitem[{\citenamefont{Shapiro}(2008)}]{Shapiro_reduced}
\bibinfo{author}{\bibfnamefont{C.}~\bibnamefont{Shapiro}}
  (\bibinfo{year}{2008}), \eprint{arXiv:0812.0769}.

\bibitem[{\citenamefont{Rudd et~al.}(2007)\citenamefont{Rudd, Zentner, and
  Kravtsov}}]{Rudd_Zentner_Kravtsov}
\bibinfo{author}{\bibfnamefont{D.~H.} \bibnamefont{Rudd}},
  \bibinfo{author}{\bibfnamefont{A.~R.} \bibnamefont{Zentner}},
  \bibnamefont{and} \bibinfo{author}{\bibfnamefont{A.~V.}
  \bibnamefont{Kravtsov}} (\bibinfo{year}{2007}), \eprint{astro-ph/0703741}.

\bibitem[{\citenamefont{Kitching
  et~al.}(2008{\natexlab{b}})\citenamefont{Kitching, Amara, Abdalla, Joachimi,
  and Refregier}}]{Kitching_formfill}
\bibinfo{author}{\bibfnamefont{T.~D.} \bibnamefont{Kitching}},
  \bibinfo{author}{\bibfnamefont{A.}~\bibnamefont{Amara}},
  \bibinfo{author}{\bibfnamefont{F.~B.} \bibnamefont{Abdalla}},
  \bibinfo{author}{\bibfnamefont{B.}~\bibnamefont{Joachimi}}, \bibnamefont{and}
  \bibinfo{author}{\bibfnamefont{A.}~\bibnamefont{Refregier}}
  (\bibinfo{year}{2008}{\natexlab{b}}), \eprint{arXiv:0812.1966}.

\bibitem[{\citenamefont{Joachimi and Schneider}(2008)}]{Joachimi_Schneider}
\bibinfo{author}{\bibfnamefont{B.}~\bibnamefont{Joachimi}} \bibnamefont{and}
  \bibinfo{author}{\bibfnamefont{P.}~\bibnamefont{Schneider}}
  (\bibinfo{year}{2008}), \eprint{arXiv:0804.2292}.

\bibitem[{\citenamefont{Bernstein}(2008)}]{Bernstein_comprehensive}
\bibinfo{author}{\bibfnamefont{G.~M.} \bibnamefont{Bernstein}}
  (\bibinfo{year}{2008}), \eprint{arXiv:0808.3400}.

\bibitem[{\citenamefont{Guzik and Bernstein}(2005)}]{Guzik_Bernstein}
\bibinfo{author}{\bibfnamefont{J.}~\bibnamefont{Guzik}} \bibnamefont{and}
  \bibinfo{author}{\bibfnamefont{G.}~\bibnamefont{Bernstein}},
  \bibinfo{journal}{Phys. Rev.} \textbf{\bibinfo{volume}{D72}},
  \bibinfo{pages}{043503} (\bibinfo{year}{2005}), \eprint{astro-ph/0507546}.

\bibitem[{\citenamefont{Heymans et~al.}(2006)}]{Heymans_STEP1}
\bibinfo{author}{\bibfnamefont{C.}~\bibnamefont{Heymans}} \bibnamefont{et~al.},
  \bibinfo{journal}{Mon. Not. Roy. Astron. Soc.}
  \textbf{\bibinfo{volume}{368}}, \bibinfo{pages}{1323} (\bibinfo{year}{2006}),
  \eprint{astro-ph/0506112}.

\bibitem[{\citenamefont{Massey et~al.}(2007)}]{Massey_STEP2}
\bibinfo{author}{\bibfnamefont{R.}~\bibnamefont{Massey}} \bibnamefont{et~al.},
  \bibinfo{journal}{Mon. Not. Roy. Astron. Soc.}
  \textbf{\bibinfo{volume}{376}}, \bibinfo{pages}{13} (\bibinfo{year}{2007}),
  \eprint{astro-ph/0608643}.

\bibitem[{\citenamefont{Paulin-Henriksson
  et~al.}(2007)\citenamefont{Paulin-Henriksson, Amara, Voigt, Refregier, and
  Bridle}}]{PaulinHenriksson_PSF}
\bibinfo{author}{\bibfnamefont{S.}~\bibnamefont{Paulin-Henriksson}},
  \bibinfo{author}{\bibfnamefont{A.}~\bibnamefont{Amara}},
  \bibinfo{author}{\bibfnamefont{L.}~\bibnamefont{Voigt}},
  \bibinfo{author}{\bibfnamefont{A.}~\bibnamefont{Refregier}},
  \bibnamefont{and} \bibinfo{author}{\bibfnamefont{S.~L.} \bibnamefont{Bridle}}
  (\bibinfo{year}{2007}), \eprint{arXiv:0711.4886}.

\bibitem[{\citenamefont{Amara and Refregier}(2007)}]{Amara_Refregier}
\bibinfo{author}{\bibfnamefont{A.}~\bibnamefont{Amara}} \bibnamefont{and}
  \bibinfo{author}{\bibfnamefont{A.}~\bibnamefont{Refregier}}
  (\bibinfo{year}{2007}), \eprint{arXiv:0710.5171}.

\bibitem[{\citenamefont{Stabenau et~al.}(2007)\citenamefont{Stabenau, Jain,
  Bernstein, and Lampton}}]{Stabenau_SNAP_sys}
\bibinfo{author}{\bibfnamefont{H.~F.} \bibnamefont{Stabenau}},
  \bibinfo{author}{\bibfnamefont{B.}~\bibnamefont{Jain}},
  \bibinfo{author}{\bibfnamefont{G.}~\bibnamefont{Bernstein}},
  \bibnamefont{and} \bibinfo{author}{\bibfnamefont{M.}~\bibnamefont{Lampton}}
  (\bibinfo{year}{2007}), \eprint{arXiv:0710.3355}.

\bibitem[{\citenamefont{Jarvis and Jain}(2004)}]{Jarvis_Jain_PSF}
\bibinfo{author}{\bibfnamefont{M.}~\bibnamefont{Jarvis}} \bibnamefont{and}
  \bibinfo{author}{\bibfnamefont{B.}~\bibnamefont{Jain}}
  (\bibinfo{year}{2004}), \eprint{astro-ph/0412234}.

\bibitem[{\citenamefont{Takada and Bridle}(2007)}]{Takada_Bridle}
\bibinfo{author}{\bibfnamefont{M.}~\bibnamefont{Takada}} \bibnamefont{and}
  \bibinfo{author}{\bibfnamefont{S.}~\bibnamefont{Bridle}},
  \bibinfo{journal}{New J. Phys.} \textbf{\bibinfo{volume}{9}},
  \bibinfo{pages}{446} (\bibinfo{year}{2007}), \eprint{arXiv:705.0163}.

\bibitem[{\citenamefont{Ma et~al.}(2008)\citenamefont{Ma, Bernstein, Weinstein,
  and Sholl}}]{Ma_jitter}
\bibinfo{author}{\bibfnamefont{Z.}~\bibnamefont{Ma}},
  \bibinfo{author}{\bibfnamefont{G.}~\bibnamefont{Bernstein}},
  \bibinfo{author}{\bibfnamefont{A.}~\bibnamefont{Weinstein}},
  \bibnamefont{and} \bibinfo{author}{\bibfnamefont{M.}~\bibnamefont{Sholl}}
  (\bibinfo{year}{2008}), \eprint{arXiv:809.2954}.

\bibitem[{\citenamefont{Heavens et~al.}(2000)\citenamefont{Heavens, Refregier,
  and Heymans}}]{Heavens_align}
\bibinfo{author}{\bibfnamefont{A.}~\bibnamefont{Heavens}},
  \bibinfo{author}{\bibfnamefont{A.}~\bibnamefont{Refregier}},
  \bibnamefont{and} \bibinfo{author}{\bibfnamefont{C.}~\bibnamefont{Heymans}},
  \bibinfo{journal}{Mon. Not. Roy. Astron. Soc.}
  \textbf{\bibinfo{volume}{319}}, \bibinfo{pages}{649} (\bibinfo{year}{2000}),
  \eprint{astro-ph/0005269}.

\bibitem[{\citenamefont{Croft and Metzler}(2000)}]{Croft_align}
\bibinfo{author}{\bibfnamefont{R.~A.~C.} \bibnamefont{Croft}} \bibnamefont{and}
  \bibinfo{author}{\bibfnamefont{C.~A.} \bibnamefont{Metzler}}
  (\bibinfo{year}{2000}), \eprint{astro-ph/0005384}.

\bibitem[{\citenamefont{Crittenden et~al.}(2001)\citenamefont{Crittenden,
  Natarajan, Pen, and Theuns}}]{Crittenden_align}
\bibinfo{author}{\bibfnamefont{R.~G.} \bibnamefont{Crittenden}},
  \bibinfo{author}{\bibfnamefont{P.}~\bibnamefont{Natarajan}},
  \bibinfo{author}{\bibfnamefont{U.-L.} \bibnamefont{Pen}}, \bibnamefont{and}
  \bibinfo{author}{\bibfnamefont{T.}~\bibnamefont{Theuns}},
  \bibinfo{journal}{Astrophys. J.} \textbf{\bibinfo{volume}{559}},
  \bibinfo{pages}{552} (\bibinfo{year}{2001}), \eprint{astro-ph/0009052}.

\bibitem[{\citenamefont{Mackey et~al.}(2002)\citenamefont{Mackey, White, and
  Kamionkowski}}]{Mackey_align}
\bibinfo{author}{\bibfnamefont{J.}~\bibnamefont{Mackey}},
  \bibinfo{author}{\bibfnamefont{M.~J.} \bibnamefont{White}}, \bibnamefont{and}
  \bibinfo{author}{\bibfnamefont{M.}~\bibnamefont{Kamionkowski}},
  \bibinfo{journal}{Mon. Not. Roy. Astron. Soc.}
  \textbf{\bibinfo{volume}{332}}, \bibinfo{pages}{788} (\bibinfo{year}{2002}),
  \eprint{astro-ph/0106364}.

\bibitem[{\citenamefont{Jing}(2002)}]{Jing_align}
\bibinfo{author}{\bibfnamefont{Y.~P.} \bibnamefont{Jing}},
  \bibinfo{journal}{Mon. Not. Roy. Astron. Soc.}
  \textbf{\bibinfo{volume}{335}}, \bibinfo{pages}{L89} (\bibinfo{year}{2002}),
  \eprint{astro-ph/0206098}.

\bibitem[{\citenamefont{Heymans and Heavens}(2003)}]{Heymans_Heavens_align}
\bibinfo{author}{\bibfnamefont{C.}~\bibnamefont{Heymans}} \bibnamefont{and}
  \bibinfo{author}{\bibfnamefont{A.}~\bibnamefont{Heavens}},
  \bibinfo{journal}{Mon. Not. Roy. Astron. Soc.}
  \textbf{\bibinfo{volume}{339}}, \bibinfo{pages}{711} (\bibinfo{year}{2003}),
  \eprint{astro-ph/0208220}.

\bibitem[{\citenamefont{King and Schneider}(2003)}]{King_Schneider_align}
\bibinfo{author}{\bibfnamefont{L.~J.} \bibnamefont{King}} \bibnamefont{and}
  \bibinfo{author}{\bibfnamefont{P.}~\bibnamefont{Schneider}},
  \bibinfo{journal}{Astron. Astrophys.} \textbf{\bibinfo{volume}{398}},
  \bibinfo{pages}{23} (\bibinfo{year}{2003}), \eprint{astro-ph/0209474}.

\bibitem[{\citenamefont{Hirata et~al.}(2004)}]{Hirata_SDSS_align}
\bibinfo{author}{\bibfnamefont{C.~M.} \bibnamefont{Hirata}}
  \bibnamefont{et~al.} (\bibinfo{collaboration}{SDSS}), \bibinfo{journal}{Mon.
  Not. Roy. Astron. Soc.} \textbf{\bibinfo{volume}{353}}, \bibinfo{pages}{529}
  (\bibinfo{year}{2004}), \eprint{astro-ph/0403255}.

\bibitem[{\citenamefont{Mandelbaum et~al.}(2006)\citenamefont{Mandelbaum,
  Hirata, Ishak, Seljak, and Brinkmann}}]{Mandelbaum_align}
\bibinfo{author}{\bibfnamefont{R.}~\bibnamefont{Mandelbaum}},
  \bibinfo{author}{\bibfnamefont{C.~M.} \bibnamefont{Hirata}},
  \bibinfo{author}{\bibfnamefont{M.}~\bibnamefont{Ishak}},
  \bibinfo{author}{\bibfnamefont{U.}~\bibnamefont{Seljak}}, \bibnamefont{and}
  \bibinfo{author}{\bibfnamefont{J.}~\bibnamefont{Brinkmann}},
  \bibinfo{journal}{Mon. Not. Roy. Astron. Soc.}
  \textbf{\bibinfo{volume}{367}}, \bibinfo{pages}{611} (\bibinfo{year}{2006}),
  \eprint{astro-ph/0509026}.

\bibitem[{\citenamefont{Bridle and Abdalla}(2007)}]{Bridle_Abdalla_align}
\bibinfo{author}{\bibfnamefont{S.}~\bibnamefont{Bridle}} \bibnamefont{and}
  \bibinfo{author}{\bibfnamefont{F.~B.} \bibnamefont{Abdalla}},
  \bibinfo{journal}{Astrophys. J.} \textbf{\bibinfo{volume}{655}},
  \bibinfo{pages}{L1} (\bibinfo{year}{2007}), \eprint{astro-ph/0608002}.

\bibitem[{\citenamefont{Bridle and King}(2007)}]{Bridle_King}
\bibinfo{author}{\bibfnamefont{S.}~\bibnamefont{Bridle}} \bibnamefont{and}
  \bibinfo{author}{\bibfnamefont{L.}~\bibnamefont{King}}, \bibinfo{journal}{New
  J. Phys.} \textbf{\bibinfo{volume}{9}}, \bibinfo{pages}{444}
  (\bibinfo{year}{2007}), \eprint{arXiv:705.0166}.

\bibitem[{\citenamefont{Hirata and Seljak}(2004)}]{HirataSeljak}
\bibinfo{author}{\bibfnamefont{C.~M.} \bibnamefont{Hirata}} \bibnamefont{and}
  \bibinfo{author}{\bibfnamefont{U.}~\bibnamefont{Seljak}},
  \bibinfo{journal}{Phys. Rev.} \textbf{\bibinfo{volume}{D70}},
  \bibinfo{pages}{063526} (\bibinfo{year}{2004}), \eprint{astro-ph/0406275}.

\bibitem[{\citenamefont{Newman}(2008)}]{Newman}
\bibinfo{author}{\bibfnamefont{J.~A.} \bibnamefont{Newman}}
  (\bibinfo{year}{2008}), \eprint{arXiv:0805.1409}.

\bibitem[{\citenamefont{Knox et~al.}(1998)\citenamefont{Knox, Scoccimarro, and
  Dodelson}}]{Knox_Scocc_Dod}
\bibinfo{author}{\bibfnamefont{L.}~\bibnamefont{Knox}},
  \bibinfo{author}{\bibfnamefont{R.}~\bibnamefont{Scoccimarro}},
  \bibnamefont{and} \bibinfo{author}{\bibfnamefont{S.}~\bibnamefont{Dodelson}},
  \bibinfo{journal}{Phys. Rev. Lett.} \textbf{\bibinfo{volume}{81}},
  \bibinfo{pages}{2004} (\bibinfo{year}{1998}), \eprint{astro-ph/9805012}.

\bibitem[{\citenamefont{Huterer and Turner}(2001)}]{Huterer_Turner}
\bibinfo{author}{\bibfnamefont{D.}~\bibnamefont{Huterer}} \bibnamefont{and}
  \bibinfo{author}{\bibfnamefont{M.~S.} \bibnamefont{Turner}},
  \bibinfo{journal}{Phys. Rev.} \textbf{\bibinfo{volume}{D64}},
  \bibinfo{pages}{123527} (\bibinfo{year}{2001}), \eprint{astro-ph/0012510}.

\bibitem[{\citenamefont{Tegmark et~al.}(1997)\citenamefont{Tegmark, Taylor, and
  Heavens}}]{TTH}
\bibinfo{author}{\bibfnamefont{M.}~\bibnamefont{Tegmark}},
  \bibinfo{author}{\bibfnamefont{A.}~\bibnamefont{Taylor}}, \bibnamefont{and}
  \bibinfo{author}{\bibfnamefont{A.}~\bibnamefont{Heavens}},
  \bibinfo{journal}{Astrophys. J.} \textbf{\bibinfo{volume}{480}},
  \bibinfo{pages}{22} (\bibinfo{year}{1997}), \eprint{astro-ph/9603021}.

\bibitem[{\citenamefont{Aldering et~al.}(2004)}]{SNAP}
\bibinfo{author}{\bibfnamefont{G.}~\bibnamefont{Aldering}} \bibnamefont{et~al.}
  (\bibinfo{collaboration}{SNAP}) (\bibinfo{year}{2004}),
  \eprint{astro-ph/0405232}.

\bibitem[{\citenamefont{{Albrecht} et~al.}(2006)\citenamefont{{Albrecht},
  {Bernstein}, {Cahn}, {Freedman}, {Hewitt}, {Hu}, {Huth}, {Kamionkowski},
  {Kolb}, {Knox} et~al.}}]{DETF}
\bibinfo{author}{\bibfnamefont{A.}~\bibnamefont{{Albrecht}}},
  \bibinfo{author}{\bibfnamefont{G.}~\bibnamefont{{Bernstein}}},
  \bibinfo{author}{\bibfnamefont{R.}~\bibnamefont{{Cahn}}},
  \bibinfo{author}{\bibfnamefont{W.~L.} \bibnamefont{{Freedman}}},
  \bibinfo{author}{\bibfnamefont{J.}~\bibnamefont{{Hewitt}}},
  \bibinfo{author}{\bibfnamefont{W.}~\bibnamefont{{Hu}}},
  \bibinfo{author}{\bibfnamefont{J.}~\bibnamefont{{Huth}}},
  \bibinfo{author}{\bibfnamefont{M.}~\bibnamefont{{Kamionkowski}}},
  \bibinfo{author}{\bibfnamefont{E.~W.} \bibnamefont{{Kolb}}},
  \bibinfo{author}{\bibfnamefont{L.}~\bibnamefont{{Knox}}},
  \bibnamefont{et~al.} (\bibinfo{year}{2006}), \eprint{astro-ph/0609591}.

\bibitem[{\citenamefont{Komatsu et~al.}(2008)}]{WMAP5}
\bibinfo{author}{\bibfnamefont{E.}~\bibnamefont{Komatsu}} \bibnamefont{et~al.}
  (\bibinfo{collaboration}{WMAP}) (\bibinfo{year}{2008}),
  \eprint{arXiv:0803.0547}.

\bibitem[{\citenamefont{Jouvel et~al.}(2009)}]{Jouvel}
\bibinfo{author}{\bibfnamefont{S.}~\bibnamefont{Jouvel}} \bibnamefont{et~al.}
  (\bibinfo{year}{2009}), \eprint{arXiv:0902.0625}.

\bibitem[{\citenamefont{{Dahlen} et~al.}(2008)\citenamefont{{Dahlen},
  {Mobasher}, {Jouvel}, {Kneib}, {Ilbert}, {Arnouts}, {Bernstein}, and
  {Rhodes}}}]{Dahlen}
\bibinfo{author}{\bibfnamefont{T.}~\bibnamefont{{Dahlen}}},
  \bibinfo{author}{\bibfnamefont{B.}~\bibnamefont{{Mobasher}}},
  \bibinfo{author}{\bibfnamefont{S.}~\bibnamefont{{Jouvel}}},
  \bibinfo{author}{\bibfnamefont{J.-P.} \bibnamefont{{Kneib}}},
  \bibinfo{author}{\bibfnamefont{O.}~\bibnamefont{{Ilbert}}},
  \bibinfo{author}{\bibfnamefont{S.}~\bibnamefont{{Arnouts}}},
  \bibinfo{author}{\bibfnamefont{G.}~\bibnamefont{{Bernstein}}},
  \bibnamefont{and} \bibinfo{author}{\bibfnamefont{J.}~\bibnamefont{{Rhodes}}},
  \bibinfo{journal}{Astrophys. J.} \textbf{\bibinfo{volume}{136}},
  \bibinfo{pages}{1361} (\bibinfo{year}{2008}), \eprint{0710.5532}.

\bibitem[{\citenamefont{Schneider et~al.}(2006)\citenamefont{Schneider, Knox,
  Zhan, and Connolly}}]{Schneider_Knox}
\bibinfo{author}{\bibfnamefont{M.}~\bibnamefont{Schneider}},
  \bibinfo{author}{\bibfnamefont{L.}~\bibnamefont{Knox}},
  \bibinfo{author}{\bibfnamefont{H.}~\bibnamefont{Zhan}}, \bibnamefont{and}
  \bibinfo{author}{\bibfnamefont{A.}~\bibnamefont{Connolly}},
  \bibinfo{journal}{Astrophys. J.} \textbf{\bibinfo{volume}{651}},
  \bibinfo{pages}{14} (\bibinfo{year}{2006}), \eprint{astro-ph/0606098}.

\end{thebibliography}
\end{document}